# A Survey on Coarse-Grained Reconfigurable Architectures From a Performance Perspective

**ARTUR PODOBAS**[1,2], **KENTARO SANO**[1], **AND SATOSHI MATSUOKA**[1,3]

[1]RIKEN Center for Computational Science, Kobe 650-0047, Japan
[2]Department of Computer Science, KTH Royal Institute of Technology, 114 28 Stockholm, Sweden
[3]Department of Mathematical and Computing Sciences, Tokyo Institute of Technology, Tokyo 152-8550, Japan

Corresponding author: Artur Podobas (artur@podobas.net)

This work was supported by the New Energy and Industrial Technology Development Organization (NEDO).

**ABSTRACT** With the end of both Dennard's scaling and Moore's law, computer users and researchers are aggressively exploring alternative forms of computing in order to continue the performance scaling that we have come to enjoy. Among the more salient and practical of the post-Moore alternatives are reconfigurable systems, with Coarse-Grained Reconfigurable Architectures (CGRAs) seemingly capable of striking a balance between performance and programmability. In this paper, we survey the landscape of CGRAs. We summarize nearly three decades of literature on the subject, with a particular focus on the premise behind the different CGRAs and how they have evolved. Next, we compile metrics of available CGRAs and analyze their performance properties in order to understand and discover knowledge gaps and opportunities for future CGRA research specialized towards High-Performance Computing (HPC). We find that there are ample opportunities for future research on CGRAs, in particular with respect to size, functionality, support for parallel programming models, and to evaluate more complex applications.

**INDEX TERMS** Coarse-grained reconfigurable architectures, CGRA, FPGA, computing trends, reconfigurable systems, high-performance computing, post-Moore.

## I. INTRODUCTION

With the end of Dennard's scaling [1] and the looming threat that even Moore's law [2] is about to end [3], computing is perhaps facing its most challenging moments. Today, computer researchers and practitioners are aggressively pursuing and exploring alternative forms of computing in order to try to fill the void that an end of Moore's law would leave behind. There are already a plethora of emerging and intrusive technologies with the promise of overcoming the limits of technology scaling, such as quantum- or neuromorphic-computing [4], [5]. However, not all *Post-Moore* architectures are intrusive, and some merely require us to step away from the comforts of von-Neumann architecture offers. Among the more salient of these technologies are reconfigurable architectures [6].

Reconfigurable architectures are systems that attempt to retain some of the silicon plasticity that an ASIC solution usually throws away. These systems – at least conceptually – allow the silicon to be malleable and its functionality

The associate editor coordinating the review of this manuscript and approving it for publication was Nitin Nitin.

dynamically configurable. A reconfigurable system can, for example, mimic a processor architecture for some time (e.g., a RISC-V core [7]), and then be changed to mimic an LTE baseband station [8]. This property of reconfigurability is highly sought after, since it can mitigate the end of Moore's law to some extent– we do not need more transistors, we just need to spatially configure the silicon to match the computation in time.

Recently, a particular branch of reconfigurable architecture – the Field-Programmable Gate Arrays (FPGAs) [9] – has experienced a surge of renewed interest for use in High-Performance Computing (HPC), and recent research has shown performance- or power-benefits for multiple applications [10]–[14]. At the same time, many of the limitations that FPGAs have, such as slow configuration times, long compilations times, and (comparably) low clock frequencies, remain unsolved. These limitations have been recognized for decades (e.g., [15]–[17]), and have driven forth a different branch of reconfigurable architecture: the Coarse-Grained Reconfigurable Architecture (CGRAs).

CGRAs trade some of the flexibility that FPGAs have to solve their limitations. A CGRA can operate at higher clock frequencies, can provide higher theoretical compute performance, can drastically reduce compilation times, and – perhaps most importantly – reduce reconfiguration time substantially. While CGRAs have traditionally been used in embedded systems (particular for media-processing), lately, they too are considered for HPC. Even traditional FPGA vendors such as Xilinx [18] and Intel [19] are creating and/or investigating to coarsen their existing reconfigurable architecture to complement other forms of computing.

In this paper, we survey the literature of CGRAs, summarizing the different architectures and systems that have been introduced over time. We complement surveys written by our peers by focusing on understanding the trends in performance that CGRAs have been experiencing, providing insights into where the community is moving, and any potential gaps in knowledge that can/should be filled.

The contributions of our work are as follows:

- A survey over three decades of Coarse-Grained Reconfigurable Architectures, summarizing existing architecture types and properties,
- A quantitative analysis over performance metrics of CGRA architectures as reported in their respective papers, and
- An analysis on trends and observations regarding CGRAs with discussion

The remaining paper is organized in the following way. Section II introduces the motivation behind CGRAs, as well as their generic design for the unfamiliar reader. Section III positions this survey against existing surveys on the topic. Section IV quantitatively summarizes each architecture that we reviewed, describing key characteristics and the premise behind each respective architecture. Section V analyzes the reviewed architecture from different perspectives (Sections VII, VIII, and VI), which we finally discuss at the end of the paper in section IX.

## II. INTRODUCTION TO CGRAs

Before summarizing the now three decades of Coarse-Grained Reconfigurable Architecture (CGRA) research, we start by describing the main aspirations and motivations behind them. To do so, we need to look at the predecessor of the CGRAs: The Field-Programmable Gate Array (FPGA).

FPGAs are devices that were developed to reduce the cost of simulation and developing Application-Specific Integrated Circuits (ASICs). Because any bug/fault that was left undiscovered post ASIC tape-out would incur a (potentially) significant economic loss, FPGAs were (and still are) crucial to digital design. In order for FPGAs to mimic any digital design, they are made to have a large degree of fine-grained reconfigurability. This fine-grained reconfigurability was achieved by building FPGAs to contain a large amount of on-chip SRAM cells called Look-Up Tables (LUTs) [20].[1] Each LUT was interfaced by a few input wires

(usually 4-6) and produced an output (and its complement) as a function of the SRAM content and their inputs. Hence, depending on the sought-after functionality to be simulated, LUTs could be configured and – through a highly reconfigurable interconnect – could be connected to each other, finally yielding the expected designs. The design would naturally run between one and two orders of magnitude lower frequency (for example, there is a $37\times$ reduction in frequency when running Intel Atom on an FPGA [21]) than the final standard-cell ASIC, but would nevertheless be an invaluable prototyping tool.

By the early 1990s, FPGAs had already found other uses (aside from digital development) within telecommunication, military, and automobile industries—the FPGA was seen as a compute device in its own right and there were some aspirations to use it for general-purpose computing and not only in the niche market of prototyping digital designs. Despite this, several limitations of FPGAs were quickly identified that prohibited coverage of a wide range of applications. For example, unlike software compilation tools that take minutes to compile applications, the FPGA Electronic Design Automation (EDA) flow took significantly longer, often requiring hours or even days of compilation time. Similarly, if the expected application could not fit a single device, the long reconfiguration overhead (the time it takes to program the FPGA) demotivated time-sharing or context-switching of its resources. Another limitation was that some important arithmetic operators did not map well to the FPGA; for example, a single integer multiplication could often consume a large fraction of the FPGA resources. Finally, FPGAs were relatively slow, running at a low clock frequency. Many of these challenges and limitations of applying FPGAs for general-purpose computing still hold to this day.

Early reconfigurable computing pioneers looked at the limitations of FPGAs and considered what would happen if one would increase the granularity at which it was programmed. By increasing the granularity, larger and more specialized units could be built, which would increase the performance (clock frequency) of the device. Also, since the larger units require less configuration state, reconfiguring the device would be significantly faster, allowing fine-grained time-sharing (multiple contexts) of the device. Finally, by coarsening the units of reconfiguration, one would include those units that map poorly on FPGAs into the fabric (e.g., multiplications), making better use of the silicon and increasing generality of the device. These new devices would later be called: Coarse-Grained Reconfigurable Architecture (CGRAs).

An example of what a CGRA looks like from the architecture perspective is shown in Figure 1. In Figure 1:a we see a mesh of reconfigurable cells (RCs) or processing elements (PEs), which is the smallest unit of reconfiguration that performs work, and it is through this mesh that a user (or compiler) decides how data flows through the system. There are multiple ways of bringing data in/out to/from the fabric. One common way is to map the device in the memory of a host processor (memory-mapped) and have the host processor orchestrate the execution. A different way is to include (generic) address generators (AGs) that can be configured to access external memory using some pattern (often corresponding to the nested loops of the application), and push the loaded data through the array. A third option is to have the reconfigurable cells do both the computation and address generation. Figure 1:b illustrates the internal of an RC element, which includes an ALU (integer and/or floating-point capable), two multiplexers (MUXs), and a local SRAM used for storage. The two multiplexers decide which of the external inputs to operate on. The inputs are usually the output of adjacent RCs, the local SRAM, a constant, or a previous output (e.g., for accumulations). The output of the ALU is similarly connected to adjacent RCs, local SRAM, or back to one of the MUXes. The operation of the RC is governed by a configuration register, here briefly shown in Figure 1:c. For simplicity, we show a single register that holds the state– however, in many architectures, each RC can hold multiple configurations that are cycled through over the application lifetime. Each of the configurations can, for example, hold the computation for a particular basic block (where live-in/out variables are stored in SRAM) or discrete kernels.

---

[1] While most FPGAs are based on SRAM LUTs, it is worth mentioning that alternatives exist, such as those (for example) built on Antifuse technology.

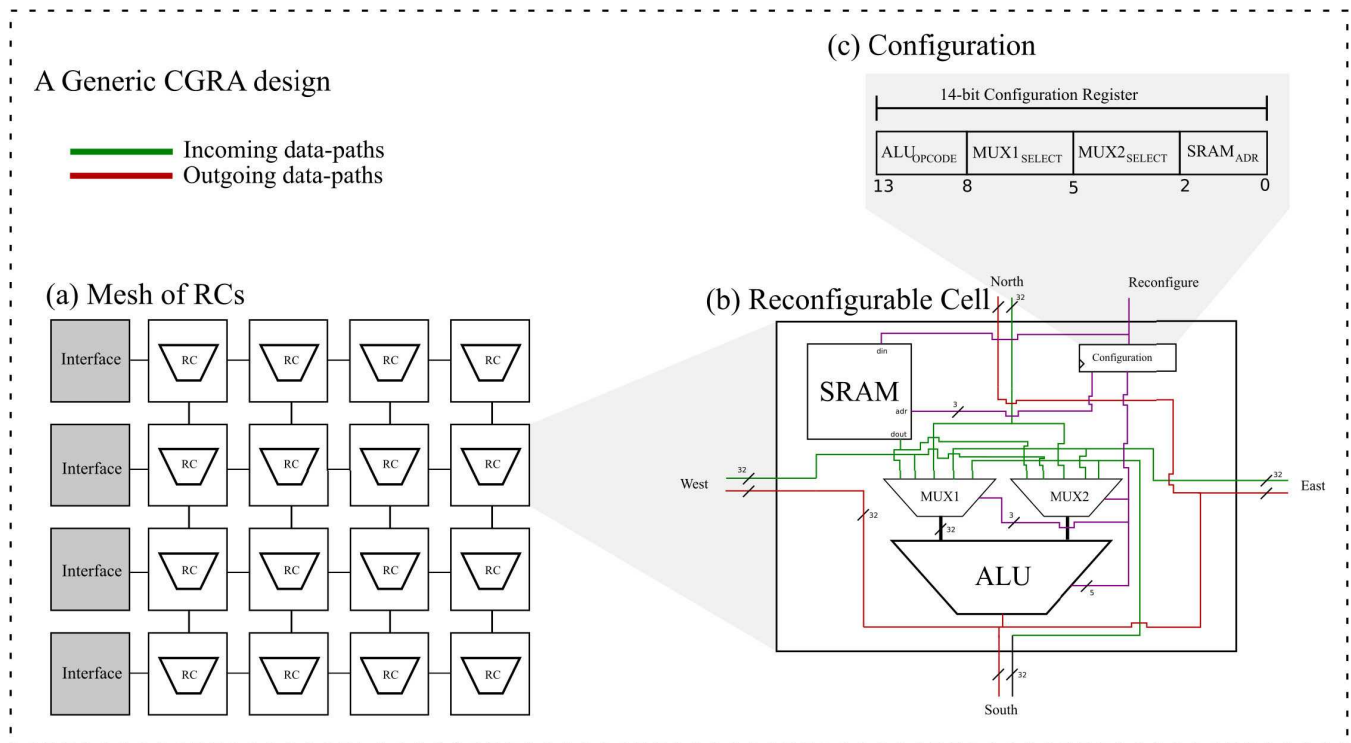

**FIGURE 1.** Illustration of a simple CGRA, showing the mesh topology (a), the internal architecture of the Reconfigurable Cell, RC (b), and an example of the configuration register (c). Although several variations exist, the illustrated structure is the predominantly used system in CGRA research.

Figure 1 illustrates how a majority of today's CGRAs look like, but at the same time, there are multiple variations. For example, early CGRAs often included fine-grained reconfigurable elements (Look-Up Tables, LUTs) inside the fabric. While the mesh topology is by far the most commonly used, some works chose a ring or linear-array topology. Finally, the flow-control of data in the network can be of varying complexity (e.g., token or tagged-token). We describe many of these in our summary in the sections that follow.

## III. SCOPE OF THE PRESENT SURVEY

Since their inception in early 1990s, CGRAs have been the subject of a plethora of research on their architecture, compilation strategies, mapping, and so on and forth. At the same time, surveys have closely monitored how the CGRA technologies have evolved through time, and we can today enjoy solid and condensed material on the subject. Surveys have covered most aspects of CGRA computing, including commercial CGRA adaptation [22], architectures [23], [24], tools and frameworks [25], and taxonomy/classification [26], [27].

The work in the present paper assumes a different position to survey the field of CGRAs. Our paper complements the existing literature by attempting to summarize and condense the performance trends of CGRA architectures, and position these against architectures such as Graphics Processing Units (GPUs) (which is what most systems use as accelerators) in order to understand what gaps future high-performance CGRA should strive to fill. To the best of our knowledge, this is the first survey providing such a comprehensive perspective within the field of CGRAs.

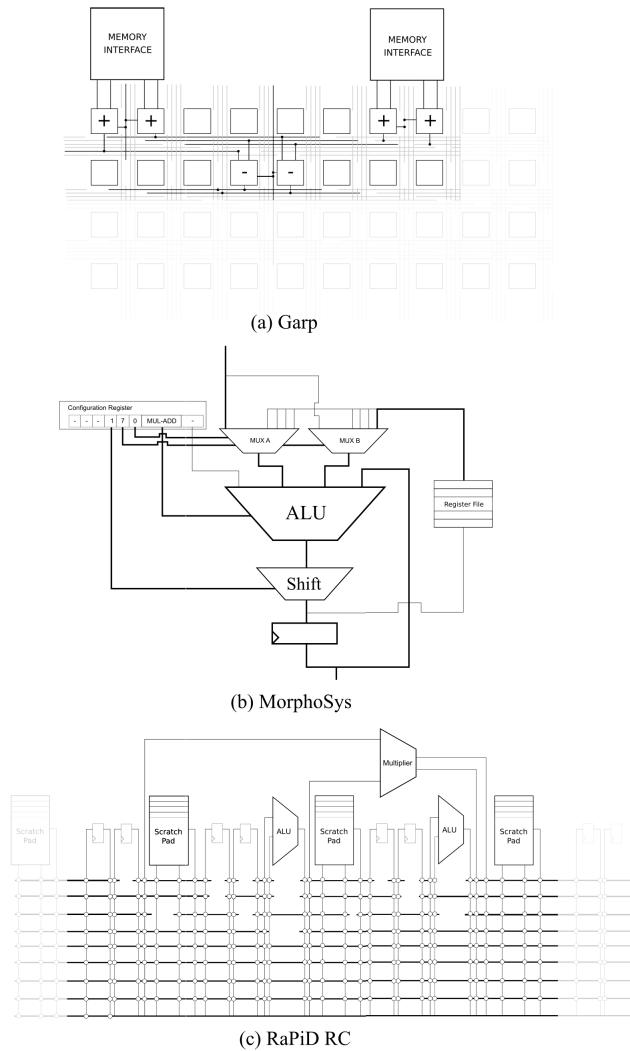

(a) Garp

(b) MorphoSys

(c) RaPiD RC

**FIGURE 2.** Three well-known early CGRA architectures that represent different approaches to the concept, where (a) Garp represents RCs with fine granularity (1-2bits), (b) MorphoSys used the structure commonly found in modern CGRAs, and (c) RaPiD adopted a linear-array of heterogeneous units that are connected through a shared segmented bus.

## IV. OVERVIEW OF CGRA ARCHITECTURE RESEARCH

### A. EARLY PIONEERING CGRAs

Some early CGRAs were not much coarser than their respective FPGAs. For example, the Garp [17] (shown in Figure 2:a) infrastructure was reconfigurable at a 2-bit (rather than FPGA 1-bit) granularity. Here, each reconfigurable unit could connect to neighbors in both the horizontal (used for carry-outs) and vertical direction, as well as to dedicated bus lines for interfacing memory. By using several reconfigurable units along the horizontal axis, users could implement arithmetic operations of varying sizes (e.g., 18-bit additions). The arithmetic units created along the horizontal directions could have their output connected along the vertical direction, creating a computational data-path. An external processor (the MIPS-based [28] TinyRISC in Garp's case) could then orchestrate the execution of this data-path. Using CGRAs as co-accelerators in this way was (and still is) a common way of leveraging them. The Garp project spanned several years and included the development of a C compiler [29].

CHESS [30] – unlike Garp – operated on a reconfigurability width of 4-bits. CHESS, as the name implies, layout the individual reconfigurable elements in a fairly uniform mesh, where elements of routing and elements of compute is altered across the mesh. Here, each reconfigurable compute element had access to all eight of its neighbors. Unlike Garp, whose reconfigurable elements were built around look-up tables (as FPGAs), CHESS used ALU-like structures with fixed functionalities that a user could choose (or configure) to use. Another interesting feature was that the compute elements could be reconfigured to act as (limited) on-chip scratchpads. D-Fabrix [31] was based on CHESS and was taped-out as a commercial product.

Raw [16], [32], [33] takes a different approach to CGRAs. Rather than keeping the reconfigurable tiles minimalistic, it instead chose to make them software programmable. Where-as Garp was based on LUTs, and CHESS was based on a single 64-bit configuration register per RC, Raw RCs have a fully dedicated instruction memory and highly dynamic network-on-chip, along with necessary hardware to support it. In fact, the Raw architecture is very similar to the modern many-core architecture, albeit lacking shared-memory support such as cache coherency. Raw spanned several years, and had a mature software infrastructure and prototype chips were taped out in 2004 [34]. It was also the precursor to the modern many-core architecture Tilera [35], which was partially built on the outcome of Raw.

The REMARC [15], [36] architecture was an early – at the time, quite coarse – architecture that operated on a 16-bit data-path. It was quite similar to modern CGRAs, since the reprogrammable elements all included an ALU, a small register file, and were directly connected to their neighbors in a mesh-like topology. Configuring the CGRA was done by programming the instruction RAM that was local to each tile with some particular functionality, where a global *program counter* (called nano-PC) synchronously orchestrated (or *sequenced*) the execution. Global communication wires ran across the horizontal and vertical axis, allowing elements to communicate with external resources. As with Garp – but unlike the Raw – the REMARC architecture was designed to work as a co-processor.

Another early but influential CGRA was the MATRIX [37] architecture, which (similar to REMARC) revolved around ALUs as the main reconfigurable compute resource, but was slightly more fine-grained than REMARC due to choosing an 8-bit (contra REMARCs 16-bit) data-path. Despite their name, the functionality of the ALU was actually more similar to that of an FPGA, where a NOR-plane could be programmed to desired functionality (similar to a Programmable Logic Array, PLAs), but did also include native support for pattern matching. The MATRIX, for its time, had a remarkably advanced network topology, where compute elements could directly communicate with neighbors on a two-square Manhattan distance. Additionally, the network included by-pass layers for remote compute elements to communicate. The network also supports computing on the data that was routed, including both shift- and reduction operations.

The MorphoSys [38], [39] (shown in Figure 2:b) was similar to the REMARC architecture, both in structure, granularity (16-bit), and also in the type of applications it targeted (media applications). MorphoSys was designed to act as a co-processor, and had the (today) well-known structure of CGRAs, which included an ALU, a small register file, an output shifter (to assist fixed-point arithmetic) and two larger multiplexers driven by the outputs of neighbors. The compute elements are arranged hierarchically in two layers: the first is a local quadrant where elements have access to all other compute elements along the vertical and horizontal axis, and the second layer are four quadrants composed into a mesh. Unlike previous CGRAs, MorphoSys had a dedicated multiplier inside the ALUs. A CGRA based on MorphoSys was also realized on silicon nearly seven years from its inception [40].

While most of the CGRAs described so-far used a mesh topology of interconnection (with some connectivity), other topologies have been considered. RaPiD [41], [42] (shown in Figure 2:c) was a CGRA that arranged its reconfigurable processing elements in a single dimension. Here, each processing element was composed of a number of primitive blocks, such as ALUs, Multipliers, scratchpads, or registers. These primitive blocks were connected to each other through a number of local, segmented, tri-stated bus lines that could be configured to form a data-path– a so-called *linear array*. These processing elements could themselves be chained together to form the final CGRA. Interestingly, RaPiD could be partially reconfigured during execution in what the authors called ''virtual execution''. RaPiD itself did not access data; instead, a number of generic address pattern generators interfaced external memory and streamed the data through the compute fabric.

The KressArray [43]–[45] was one of the earliest CGRA designs to be created, and the project spanned nearly a decade with multiple versions and variants of the architecture. It features a hierarchical topology, where the lowest tier was composed of a mesh of processing elements. The processing elements interfaced with neighbors and also included predication signals (to map if-then-else primitives). Generic address generators supported the CGRA fabric by continuously streaming data to the architecture.

Chimaera [46] was a co-processor conceptually similar to Garp, with an array of reconfigurable processing elements operating at quite a fine granularity (similar to modern FPGAs) that could be reconfigured to perform a particular operation. It was closely coupled to the host processor to the point where the register file was (in part) shadowed and shared. Mapping application to the architecture was assisted by a ''simple'' C compiler, and they demonstrated performance on Mediabench benchmarks [47] and the Honeywell ACS suite [48].

PipeRench [49] applied a novel network topology that was a hybrid between that of a mesh and a linear array. Here, a large number of linear arrays were layered, where each layer sent data uni-directionally to the next layer. Several future CGRAs would adopt this kind of structure, including data-flow machines (e.g., Tartan) and loop-accelerators (e.g., FPCA). The layers themselves in PipeRench were fairly fine-grained and comparable to Garp as they had reconfigurable Look-Up Tables rather than fixed-function ALUs within. PipeRench introduced a virtualization technique that treated each separate layer as a discrete accelerator, where a partial reconfiguration traveled alongside with its associated data, reconfiguring the next layer according to its functionality in a pipeline fashion, which was new at the time. PipeRench was also later implemented on silicon [50].

The DReAM [51] architecture was explicitly designed to target the (then) next-generation 3G networks, and argues that CGRAs are well suited for the upcoming standard with respect to software-defined radio and the flexibility to hot-fix bugs (through patches) and firmware. The system has a hierarchy of configuration managers and a mesh of simple, ALU-based, RCs operating on 16-bit operands and with limited support for complex operations such as multiplications (since operations were realized through Look-Up Tables).

So far, all architectures reviewed have been computing using integer arithmetics. Imagine [52] was among the early architectures that included hardware floating-point arithmetic units. The architecture itself was similar to RaPiD—it was a linear array, where each processing element had a number of resources (scratchpads, ALUs, etc.) all connected using a global bus. Similar to RaPiD, the processing elements were passive, and external drivers were responsible for streaming data through the processing elements. The Imagine architecture had a prototype realized six years after its seminal paper [53].

## B. MODERN COARSE-GRAINED RECONFIGURABLE ARCHITECTURES

Most modern CGRA architectures' lineage can be linked back to those described in the previous section, and a majority of these architectures follow the generic template that was described in the previous section. However, while the overall template remains similar, many recent architectures specialize themselves towards a certain niche use (low-power, deep learning, GPU-like programmable, etc.).

The ADRES CGRA system [54], [55] (Figure 3:a) has been a remarkably successful architecture template for embedded architectures, and is still widely used. ADRES is a template-based architecture, and while the most common example arranges RC's in a mesh, users are capable of defining arbitrary connectivity. Inside each element, we find an ALU of varying capability and a register file, alongside the multiplexers configured to bring in data from neighbors. The first row in the mesh, however, is unique, as an optional processor can extend its pipeline to support interfacing that very first row in a Very Long Instruction Word [56] (VLIW) fashion. ADRES, by design, is thus heterogeneous. ADRES comes with a compiler called DRESC [57], which can handle the freedom that ADRES allows with respect to arbitrary connectivity. ADRES as an architecture has been (and still is) a popular platform for CGRA research, such as when exploring multi-threaded CGRA support [58], topologies [59], asynchronous further-than-neighbor communication(e.g. HyCUBE [60]), or CGRA designs frameworks/generators (e.g. CGRA-ME [61], [62]). Furthermore, ADRES has been taped out on silicon, for example in the Samsung Reconfigurable Processor (SRP) and the follow-up UL-SRP [63] architecture.

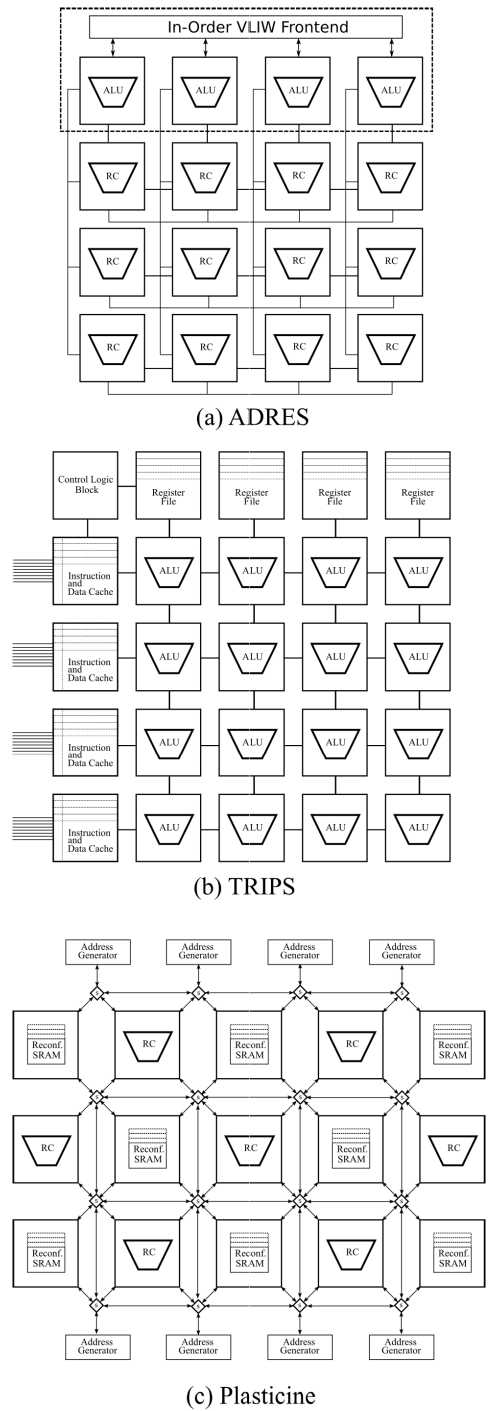

(a) ADRES

(b) TRIPS

(c) Plasticine

**FIGURE 3.** The (a) ADRES architecture was a CGRA template architecture that also later was commercialized (by among others Samsung); unique to ADRES was that the first row of RCs extended the backend pipeline of the VLIW processor that orchestrated the execution. (b) The TRIPS architecture was among the first to replace the traditional super-scalar processor pipeline with a relatively large CGRA-like mesh in order to exploit more parallelism. (c) The Plasticine architecture is a recent CGRA architecture that focuses on parallel patterns through specialized pattern address generators (for both external and internal storage).

The Dynamically Reconfigurable ALU Array (DRAA) [64] was a generic CGRA template proposed in 2003 to encourage compilation research on CGRA architecture. Architecture-wise, DRAA allowed changing many of the parameters that define a CGRA, such as the data-path width, the interconnect, size of the register file, etc. Preceding both DySER and ADRES, DRAA as a template has been used to e.g. study the memory hierarchy of CGRAs [65].

The TRIPS/EDGE [66], [67] microarchitecture (Figure 3:b), was a long-running influential project that attempted to move away from the traditional approach of exploiting instruction-level parallelism in modern processors. The premise behind TRIPS was that as technology reduced the sizes of transistors, wire delays and paths would dominate latency, and that it would be hard to continue to scale the communication in superscalar processors [68]. Instead, by tightly coupling functional units in (for example) a mesh, direct neighbor communication could easily be scaled. In effect, TRIPS/EDGE *replaced* the traditional superscalar Out-of-Order pipeline with a large CGRA array: single instructions were no longer scheduled, but instead, a new compiler [69], [70] was developed that scheduled entire blocks (essentially CGRA configurations) temporally on the processor, allowing up to 16 instructions to be executed at a single time (and many more in-flight). The TRIPS architecture was taped out on silicon [71], [72], and – despite being discontinued – represented a milestone of true high-performance computing with CGRAs. An interesting observation, albeit not necessarily related to CGRAs, is that the Edge ISA has recently received renewed interest as an alternative to express large amounts of ILP in FPGA soft processors [73].

The DySER [74] architecture integrates a CGRA into the backend of a processor's pipeline to *complement* (unlike TRIPS that replace) the functionality of the traditional (super-)scalar pipeline and has been integrated in the OpenSPARC [75] platform [76]. The key premise behind DySER is that there are many local hot regions in program code, and higher performance can be obtained by specializing in accelerating these inside the CPU. DySER was evaluated using both simulator-based (M5 [77]) and an FPGA implementation on well-known benchmarks (PARSEC [78] and SPECint) and compared with both CPU and GPU approaches, showing between 1.5×-15× improvements over SSE and comparable flexibility and performance to GPUs. Recently (2016 onwards), DySER has been the focus of the FPGA-overlay scene (see Section IV-G). Other similar work to DySER that integrates CGRA-like structures into processing cores with various goals includes CReAMS/HARTMP [79], [80] (applies dynamic binary translation) or CGRA-sharing [81] (conceptually similar to what AMD Bulldozer architecture [82] and UltraSPARC T1/T2 did with their floating-point units).

The AMIDAR [83] is another exciting long-running project that (amongst others) uses CGRA to accelerate performance-critical sections. The AMIDAR CGRA extends the traditional CGRA PE architecture with a direct interface to memory (through DMA). There is support for multiple contexts and hardware support for branching (through dedicated condition-boxes operating on predication signals), which also allows speculation. The AMIDA CGRA has been implemented and verified on an FPGA platform, and early results show that it can reach over 1 GHz of clock frequency when mapped to a 45 nm technology.

The Versat CGRA [84] is a co-processor system that primarily targets different transforms (IDCT, FFT, etc.) and filters (FIR, IIR, etc.), and the authors argue that for these kernels, there is no need to use unnecessarily large CGRAs, and that minimal power-efficient CGRAs are preferred. The Versat CGRA is a linear array, where a number of functional units share a data-bus, whose output connects back to the inputs of the functional units. It is programmed using a C/C++ dialect. The Versat was evaluated against an ARM Cortex A9, and experienced between $1.7\times$ (IIR) and $17.7\times$ (FFT) performance increase over said processor.

The MORA [85] architecture was a platform for CGRA-related research. MORA targeted media-processing, and hence provided an 8-bit architecture with processing elements covering the most commonly used operations. MORA itself was similar to the previous MATRIX, with a simple 2D mesh structure with neighbor communication. Each processing element had a scratchpad (256 bytes large). MORA was programmable using a domain-specific language developed over C++ [86].

The CGRA Express [87] is yet another architecture that follows the concept of being a mesh with simple, ALU-like structures. The premise and motivation for their work is that most existing CGRA applications are optimized for maximal graph coverage rather than sequential speed. The hypothesis is that – depending on the operators each PEs is configured to use – they can exploit the resulting positive clock slack of the operators and cascade (fuse) more operations per clock cycle than blindly registering the intermediate output. This, in turn, allows them to execute more instructions per cycle (or reduces the frequency) with little performance losses. In their architecture, they add an extra bypass network that can be configured to not be pipelined. They show both power and performance benefits on multimedia benchmarks with and without their approach. The work could be conceptually seen as the opposite to what modern FPGAs (e.g., Stratix 10) do with Hyperflex [88], but for CGRAs.

The polymorphic pipeline array (PPA) [89] performed an interesting pilot study that drove the parameters of their CGRA: they simulated a large number of benchmarks scheduled on a hypothetical (infinite) CGRA, with focus on modulo-scheduling and loop unrolling. They revealed that even with infinitely large CGRAs, the performance levels will be bound as a function of the instruction-level parallelism in the loops and the limitation of modulo-scheduling, and they argue that there is a definitive need to include other forms of parallelism to scale on CGRAs. While the PEs themselves follow a standard layout, they propose an interesting technique that allows multiple (unique) kernels to be executed concurrently on the CGRA, where each kernel communicates with each other either through DMA or shared memory. Kernels can also be resized to fully exploit the CGRA array.

The premise behind SIMD RA [90] is similar to that of PPA: CGRAs relies too much on instructional-level parallelism, and opportunities from other forms of parallelism are lost. SIMD RA focuses on embedding support to modularize the CGRA-array to supporting multiple discrete controllable regions that (may) operate in SIMD fashion. They found that using SIMD not only yielded better performance, but was also more area-efficient compared to using software pipelining only.

SmartCell [91] was a CGRA that aspired to be low-power and provide high-performance, supporting both SIMD- and MIMD-type parallelism. The architecture was effectively a 2D mesh, but with the mesh divided into $2 \times 2$ quadrants of processing elements. These $2 \times 2$ islands shared a reconfigurable router, and inter-quadrant communication was limited to the connectivity of these routers. The processing elements themselves were fairly standard and contained instruction memory whose instruction (configuration) was set either per processing element (MIMD) or sequenced globally (SIMD).

The Butter [92] architecture was an early effort at providing a CGRA template with subword compute capabilities, meaning that while the data-path width was 32-bit it could be segmented into carrying multiple values and operate on them in SIMD fashion. The size was fixed to $4 \times 8$ PEs (changed in the later SCREMA architecture, see section IV-G), had multiple contexts, and the CGRA is programmed using C functions. The Butter architecture was synthesized towards both ASICs and FPGAs, while its sequels (CREMA and SCREMA) were more FPGA-focused (described in later section IV-G).

BilRC [93] is a heterogeneous mesh composed of three different blocks: generic ALU blocks, Multiplication/Shifter nodes, and memory blocks, following the (by now) traditional recipe of a CGRA. Unique to BilRC is that the architecture explicitly exposes the triggering of instructions, allowing the programmer and/or application fine-grained control over the amount of parallelism or when instructions are executed.

The lack of floating-point support in CGRAs has also been a driving force for research. FloRA [94] is 16-bit IEEE-754 floating-point capable CGRA. The architecture itself is composed of 64 RCs, and each RC is fairly standard and does not include a dedicated floating-point core itself; instead, two RCs can be combined to enable single-precision (32-bit) floating-point support, where mantissa and exponent-computation is distributed among the pair.

Feng *et al.* [95] introduce a floating-point capable architecture specifically designed for radar signal processing. Despite the familiar mesh-based interconnection, the design deviates from the traditional approach since their processing elements are fairly diverse and heterogeneous. The CGRA itself was taped out on silicon and could reach up to 70 GFLOP/s floating-point performance.

The recent TRANSPIRE architecture [96] is a small $2 \times 4$ CGRA that targets near-sensor Internet of Things (IoT) use cases. Realizing the importance of variable precision floating-point computations, the TRANSPIRE architecture supports both the IEEE-754 single-precision computations and a recently proposed 8-bit format called binary8 [97]. When configured in binary8, the architecture is capable of leveraging SIMD parallelism to increase performance and energy efficiency. The CGRA was empirically shown to perform up to ten times better than a RISC-V [98] processor with a comparable silicon footprint and could reach up to 224 GOPs/W of binary8 performance and 156 GOPs/Watt when employing full 32-bit precision for selected near-sensor kernels.

A PRET-driven (Precision Timed) CGRA aimed towards predictable real-time processing was developed by Siqueira and Kreutz [99]. Interestingly, the CGRA has support for threads, which is a concept used more in high-performance designs rather than in real-time designs. The architecture is similar to a SIMT (Simultaneous Multi-Threading) architecture, where each processing element has a duplicate number of resources (primarily the register files) that are unique to each thread. Aside from having deterministic timing inside the CGRA, the authors also implemented a predictable external memory access, required for real-time systems.

The recent Sparse Processing Unit (SPU) [100] architecture aspires to provide a CGRA for general-purpose computing. The main novelty is that SPU extends existing CGRA designs with support for two types of computational patterns: what they call "stream-joins" (e.g., sparse vector multiplication inner-product) and alias-free scatter/gather (regular loops with indirection). This is achieved by extending the typical CGRA with options to conditionally consume input tokens (re-use values), reset accumulators, or conditionally discard output tokens. Address generation units (linear and non-linear) reside inside on-chip SRAM controllers. The SPU targets general-purpose workloads with some favor towards deep learning applications.

The premise of Softbrain [101] is to combine both vector-level (for regular, efficient memory-access) and data-flow (for instruction-level parallelism) computation in CGRAs to reach high performance and power-efficiency. The architecture consists of a number of stream-engines (essentially address generators in prior work) and the CGRA substrate itself. The input to the CGRA substrate is a number of vector ports (512-bit memory interfaces), the on-chip scratchpad, or the local output feedback, and a stream-controller that orchestrates the execution of the system.

The Chameleon [102] CGRA was an early commercial CGRA that focused on competing with early DSPs and FPGAs. Here the CGRA is layered, where they call each layer a slice. Each slice has three tiles, where each tile has a number of scratchpad memories that interface with eight processing elements that can be reconfigured. The idea is to load the local scratchpad with data, configure the processing elements associated with the scratchpad with some functionality, and then pipe the data onto other slices. The Chameleon was implemented in a $0.25um$ process running at a 125 MHz clock frequency. The architecture itself operates on 32-bit data-path width but can be configured to divide the data stream into two 16-bit or four 8-bit streams as well.

SiLago (Silicon Large Grain Objects) [103] is a methodology for creating CGRA-based platforms. The premise behind the method is to use reconfigurable CGRA processing elements (based on DRRA [104]) as building blocks for future systems in order to reduce production cost with little impact on performance (compared to hand-made ASICs). Platforms based on SiLago and DRRA are (among others) specialized for deep learning [105], Brain-simulation computing [106], and genome identification [107]. The Q100 [108] is similar in concepts but specializes in data-base computing and provides tiles for computing on data-flow streams that users can assemble larger systems from.

## C. LARGER CGRAs

Most CGRA systems (e.g., ADRES, TRIPS, DySER) limit the size of the array to less than 64 processing elements, and only a few of so-far mentioned CGRAs are larger than that (PipeRench had 256 PEs, Garp had 768) but they are relatively fine-grained. Likely the limited size of these CGRAs was due to their application domain, which mostly involved image-, audio-, or telecommunication applications. However, in recent years, even larger, more powerful, CGRA-based systems have emerged, many of which explicitly target High-Performance Computing.

The eXtreme Processing Platform [109] (XPP) was a CGRA that focused on multiple levels of parallelism, including that of pipeline processing, data-flow computing, and task-level execution. XPP's interconnection was deep and complex, consisting of multiple levels of various functionality. At the lowest tier, small processing elements containing scratchpad, an ALU, and associated configuration manager reside in mesh-like connectivity called a cluster. These clusters themselves are connected through switch-boxes running along their vertical and horizontal axes. Each tier had a configuration manager that is responsible for the functionality of that layer (and below), allowing fine-grained control and partitioning of the functionality of the system. XPP was token-driven, and execution of operation occurs only when data is present at inputs.

The High-Performance Reconfigurable Processor [110] (HiPReP) is an on-going CGRA research platform capable

of floating-point computations. HiPReP has dedicated floating-point circuitry (unlike, e.g., FloRA). Processing elements are organized in a mesh with a heterogeneous (in terms of bandwidth) interconnect, and include address generation units for driving data through the device. The HiPReP explicitly targets high-performance computing.

WaveScalar [111]–[113] was an exciting architecture that focused on whole-application mapping onto token-driven, data-flow CGRA-like architecture. Most CGRA systems limit the size of the CGRA fabric to the point where basic-blocks (instructions without branches) have to be split temporally across the fabric. WaveScalar takes a different approach: do not limit the CGRA fabric; instead, create a large fabric and remove all materialization of control-flow in the application by converting them to data-dependencies, enabling them to map to the architecture. The WaveScalar architecture is larger, with over 500 RCs, and – as with its rival TRIPS – was evaluated on a subset of SPECint2000, SPECfp2000, and SPEC2000 benchmarks.

Tartan [114] was another hierarchical architecture. It was a mix between a traditional mesh-based topology (e.g., ADRES) and a layered linear array (e.g., PipeRench). The system was fine-grained and operated on an 8-bit data-path with primarily addition and logical operations (no native multiplication). However, the architecture was not easily reconfigured, and expected the full application (or the majority of it) to be fully placed onto the architecture to prevent any form of context-switching. Tartan itself represents the extreme case where the majority of silicon is spent on simple compute elements, and was evaluated using SPECint [115]. The EGRA [116] architecture was similar in topology to Tartan, and had a mesh where each node consists of layered programmable ALUs. Both Tartan and EGRA were evaluated primarily using software simulators (SimpleScalar [117] and in-house simulator for Tartan).

While most CGRAs target a classic SISD architectural model – with a few exceptions for SIMD – the SGMF [118], [119] architecture instead focuses on the SIMT [120] (Single-Instruction Multiple-Thread) model that is commonly found in GPUs and programmed using languages such as CUDA [121] or OpenCL [122]. The SGMF is claimed to be similar in area to the Nvidia Fermi [123] or Kepler architecture. The architecture itself is a mesh-like CGRA with thread-tagged token flow-control and with support for synchronization. Although there are limits on what CUDA constructs can be mapped (e.g., atomic operations are not supported), the architecture itself is shown through simulation to be a viable and competitive alternative to existing GPUs.

REMUS [124] is a relatively large – for embedded SoC standards – CGRA with 512 processing elements that are driven by two ARM processors. REMUS is fairly standard in terms of layout and uses a layered mesh, although with extra temporary registers capable of holding state along the horizontal lines of the mesh, increasing opportunities for routing and more aggressive pipelining (to increase operating frequency).

The Dynamically Reconfigurable Processor [125] (DRP) was a CGRA that targeted stream-based computation. The processing elements were divided into eight tiles, each containing $8 \times 8$ processing elements, where each processing element had an ALU, input selectors (multiplexers), and an instruction memory (for multiple contexts). A number of scratchpad memories sat at the fringes of each tile and were used to store streamed data. The operation of the processing elements was controlled by an instruction pointer that was governed by hierarchical sequencers (one per tile and one global). The sequencer – effectively a programmable FSM – dictated which context was being executed, and could (re)act on signals from the tiles themselves. DRP was commercially taped out in the DRP-1 prototype by NEC.

The commercial DAPDNA-2 [126] processor produced by IPFlex contained up to 376 32-bit processing elements, organized as $8 \times 8$ PE quadrants in a mesh. The architecture was heterogeneous, with discrete tiles supporting ALU operations, scratchpad, programmable delay lines, simple address generators (counters), and I/O buffers. The processing elements contained both multiplication and arithmetic units and also supported optional pre-processing of inputs through rotation/masking units. The tiles were interconnected using horizontal and vertical busses that ran in-between and through the mesh, and crossing the quadrants could only be done at border tile locations. Performance of selected applications (FIR, FFTs, Image processing) showed two orders of magnitude better performance over the then state-of-the-art Pentium 4 processor.

The 167-processor architecture [127] was architecturally similar to both a CGRA and a conventional multi-core processor, and we include it here since the processing elements are simple, and communication between them is only performed using direct (yet dynamically configured) connectivity (and not through shared-memory or cache coherence as done in multi-core). The main focus of this work is to reduce power consumption through a series of advanced low-level optimizations (DVFS, clock generation and distribution, GALS [128], etc.). They show performance of up to 196.7 GMAs/Watt when fabricated in 65 nm technology. Other similar architectures, based on programmable cores with limited connectivity, were the IMAPCAR [129]/IMAP-CE CGRA [130] from NEC aimed towards image recognition in automobiles.

The RHyMe/REDEFINE [131], [132] architecture is a CGRA targeting High-Performance Computing (HPC) kernels. It follows a fairly typical CGRA design, where processing elements are connected through in a torus network. The premise of their work is that there is a need to exploit multiple levels of parallelism (instruction-, loop- and task-level parallelism), albeit the current implementation focuses primarily on instruction-level parallelism through modulo-scheduling. The Rhyme-Redefine supports floating-point computations.

Plasticine [133] (Figure 3:c) is a recent, large CGRA that focuses on parallel patterns. At the highest abstraction layer, it is built of a mesh of units. There are two types of units:

compute and memory units, both of which are programmable with patterns. Inside the compute units, we find the raw functional units (the ALUs) as well as a programmable state for controlling them. The compute units are built with both SISD- and SIMD-type parallelism in mind, and vector operations map natively to these units. Similarly, inside the memory units, we find a small set of ALUs coupled with programmable logic to interface the SRAM local to the units. The mesh itself interfaces external memory through a set of address generators and coalescing units. More importantly, Plasticine targets floating-point intensive applications, which is also shown in their evaluation (only three out of 13 applications are integer only). Plasticine is programmable using Spatial [134]– a custom language based on patterns for data-flow computing.

The Riken High-Performance CGRA [135] (RHP-CGRA) is a recent architecture template that targets the exploration of CGRAs for use in HPC environments. Several architectural parameters can be modified (e.g., data-path width, size in RCs, functionality, SIMD width), and the system supports co-integration with third-party memory simulators (DRAM-Sim3 [136]), allowing future memory technology to be evaluated with CGRAs. The system is token-based, supports sub-word parallelism, and leverages address generators to bring data in and out. An example design (9 × 9 mesh with 7 × 9 RCs) of the system was evaluated with several different memory technologies (DDR3, GDDR6, and HBM2) and was capable of reaching between 44.5 GOP/s (111.4 OP/cycle) and 177.2 GOP/s (234 OP/cycle) on a variety of benchmarks.

Recently, the Cerebras Wafer Scale Engine [137] has been created explicitly for high-end deep learning training. Little information is publicly available, but the architecture seems to contain both software programmable tiles and specialized tiles for tensor computations, which could make it the single largest CGRA to date with a size of over $46,225\ mm^2$.

### D. LINEAR-ARRAYS AND LOOP-ACCELERATORS

VEAL [138] was a linear array that explicitly targets the acceleration of small, compute-intensive loop-bodies. Similar to before-mentioned PPA, the authors behind VEAL performed a rigid empirical evaluation of the benchmarks they target, and demonstrate that one of the main limitations to the performance of mapping said benchmarks to CGRA fabrics is not the number of resources, but actually the amount of instruction-level parallelism extracted by modulo-scheduling. The VEAL linear array was fed by a number of custom address generators, which broadly corresponds to the induction variables of the loops that were executed. An interesting observation is that VEAL was among the few CGRA works that use double-precision arithmetics. Another loop-accelerator similar to VEAL was the FPCA [139].

The BERET [140] architecture was yet another linear array that was designed to accelerate hot regions of code. One of BERETs main contributions was the identification of a small set of graphs that the processing elements should cover (called SEBs); the set was empirically extracted from the benchmark and has since then been used in other studies (e.g., SEED [141], which is similar but improved in concept).

### E. DEEP LEARNING CGRAs

Deep learning [142], in particular the computationally regular Convolutional Neural Networks (CNNs), have lately become a target for specialized CGRAs. Here the focus is to limit the generality and reconfigurability of traditional CGRA to fit the computational patterns of CNNs and instead spend the gained logic on supporting specialized operations for the intended deep learning workloads (such as compression, multicasting, etc.). Furthermore, these architectures often honor smaller (or mixed) number representations, since deep learning often is amendable to lower-precision calculations [143].

The DT-CGRA [144], [145] architecture follows a CGRA design with relatively coarse processing elements that include up to three multiply-accumulate instructions. The processing elements also include programmable delay lines to easier map temporally close data. Data inside the RCs is synchronized through tokens. Support for multiple common deep learning access patterns (with stride, type, etc.) is facilitated through custom stream-buffers units that are programmable in a VLIW-fashion, and that generate accesses to external memory.

The Sparse CNN (SCNN) [146] is a deep learning architecture that primarily targets CNNs and can exploit sparseness in both activations and kernel weights. The architecture itself is composed of a mesh of RCs, where each element also includes a 4 × 4 multiplier array and a bank for accumulation registers. These RCs are driven- and orchestrated-by a layer sequencer, which fetches and broadcasts (compressed) weights and activations. SCNN supports inter-PE parallelism in the form of spatial blocking/tiling, where each block is artificially enlarged with a halo region, which is exchanged between adjacent tiles at the end of the computation. They also implement a *dense* version (DCNN) of the architecture in order to measure the area overhead and power- and performance-gains of including sparsity in the accelerator.

Liang *et al.* [147] introduce a CGRA accelerator that targets *reinforced learning*. The processing elements themselves are fairly static, with support for addition, multiplication, or a fusion of both. Additionally, a number of different activation functions (ReLu, sigmoid, and tanh) can be selected using the configuration register, and data can be temporally stored in a local scratchpad. Unlike most current CGRAs that place address generators in discrete units outside the RCs, Liang *et al.*'s RC include address generators inside. Global communication lines allow the user to control the reinforced training experience of the system.

The Eyeriss deep learning inference engine [148], [149] follows a CGRA design methodology as well, albeit with more focus on re-configuring the network access patterns rather than the compute (which mostly is based on multiply-accumulate operations). The CGRA itself is a mesh with a variety of options of point-to-point and broadcast operations, highly suitable for deep learning convolution patterns.

Additionally, the platform supports compression of data and exploits sparseness of intermediate activations to increase observed bandwidth. The Eyeriss architecture, depending on the type of neural network used, can utilize nearly 100% of the CGRA resources when inferring AlexNET.

One of the most recent (and perhaps radical) changes to the FPGAs is coming in the form of support for deep learning CGRAs. The Xilinx Versal [150], [151] series allocates a large part of the silicon to a mesh-like CGRA structure of programmable, neighbor-communicating, processing elements. The elements themselves are fairly general-purpose, but are marketed as targeting deep learning and telecommunication application. To remedy the eventuality that the AI engine might be missing crucial parts of deep learning functionality that has yet to come, the AI engine can directly interface remaining parts of the reconfigurable (FPGA) silicon, which is in the form of the fine-grained reconfigurable cells that Xilinx is known for. The system itself is an attempt to combine the best of both the fine-grained and coarse-grained reconfigurable worlds.

### F. LOW-POWER CGRAs
CGRAs have also been shown to be competitive in terms of power consumption, particularly when compared to existing (low-power) processors and DSP engines. The CGRAs in this domain follow the same concept as earlier CGRA designs, but focus on both technology and architecture improvements to reduce the static and/or dynamic power of the fabric.

These CGRAs tend to focus on reducing the frequency and voltage as much as possible. Since the dynamic power consumption of a system is a function of both frequency and voltage ($P_{dynamic} = C * V^2 * f_{clk}$), reducing frequency can have a dramatic effect on power consumption. Several CGRAs in this area operate on near-MHz levels, and some even remove the clock altogether.

The Cool Mega-Array [152], [153] (CMA-1 and CMA-2) architecture builds on the following two premises: (**i**) the clock (clock-tree, flip-flops, state, etc.) is the culprit behind much of the consumed power on a modern chip, and (**ii**) applications have adequate parallelism to freely trade silicon for performance where needed. The CMA-1 is a typical CGRA mesh architecture, but created without a single clock. The architecture focuses on stream-computing, where a processor presents inputs to the CGRA that – in due time – are computed using the clock-less fabric. The architecture (and its follow up, CMA-2) is power-efficient, and experiments on taped-out versions showed that the leakage power of the chip could be as low as 1 mW. The CMA architecture manages to reach up to 89.28 GOPS/Watt using a 24-bit data-path. The CMA architecture is still being researched, and recent work has focused on improving performance (through variable-latency pipelines in VPCMA [154] or further reducing power consumption through body-biasing.

The SYSCORE [155] architecture is another similar architecture that focuses on low-power consumption, but leverages dynamic scaling of both voltage and frequency (DVFS) for power-benefits, and uses a fixed-point (and not floating-point) number representation. As with CMA-1/2, it is a 24-bit data-path with a standard mesh-like arrangement of CGRA-tiles.

The i-DPs [156] architecture is an embedded biosignal processing CGRA that allows multiple configurations to share the CGRA substrate through reconfiguration. The main contribution of i-DPs is an extension of the configuration controller, which can manage RC allocation to different processors. As with most CGRAs in the biosignal domain, the i-DP runs at a low 2 MHz clock frequency, and shows both runtime and energy efficiency gains when running ECG applications.

Lopes *et al.* [157] evaluated a standard mesh-like CGRA for use in real-time biosignal processing. The CGRA they constructed had the additional benefit of being able to power-down sections of the CGRA when unused to extend battery life. Another biomedical CGRA was introduced by Duch *et al.* [158], and uses a mesh-like composition and a *1* MHz clock-frequency to accelerate electrocardiogram (ECG) analysis kernels.

The Samsung UL-SRP [63] was designed for biomedical applications. The UL-SRP is based on the ADRES, and features a hybrid high-power/high-performance mode as well as a low-power/low-performance mode which covers the different needs and scenarios.

The PULP [159] cluster system features a 16 RC mesh to improve performance and energy-consumption for near-sensor data analytics. The CGRA (called IPA [160]) is standard in the design and adopts most concepts that we have described so far, with the RCs connected in a torus fashion. The RCs are capable of 32-bit operations and feature a discrete power-controller (implementing clock-gating) for reducing energy usage when idle, and the array is capable of running at 100 MHz targeting various image processing kernels.

A different – yet equally interesting – form of power reduction is to extend the CGRA RCs to support approximate computing. X-CGRA [161] is one such example showing that adapting approximate computing in kernels that are resilient to errors (e.g., image manipulation) can drastically reduce power consumption by up to 3.21× compared to exact-methods, with as little as 4% loss in quality.

### G. OVERLAYS: CGRAs ON-TOP OF FPGAs
Some of the original incentives for incepting CGRAs was to build faster – still reconfigurable – hardware accelerators, citing FPGAs as inadequate with respect to programmability, performance, and reconfiguration overhead. While FPGA vendors did indeed remedy parts of these problems (even including floating-point DSPs [162] into the fabric), problems with compilation times and reprogramming overhead still persist to this day. By the early 2010s, several research groups had started to experiment on encapsulating the typical fine-grained resources on FPGAs with CGRA-like structures in the hope of including some of their benefits; these architectures came to known as FPGA overlays.

Overlays remedy two large performance and usability problems with FPGAs: compilation times and reconfiguration overheads. It is well-known that compiling a design towards FPGAs is a time-consuming task. Modern FPGAs are larger with many unique characteristics making it non-trivial to place and route on them. Furthermore, as the FPGAs grow, so does also the memory footprint of the synthesis tools, and it not uncommon for tools compiling against new devices (e.g., Intel Stratix 10) to consume a large portion of the system memory, effectively restricting the number of parallel compilations possible in a machine. By coarsening and reducing the number of reconfigurable units, compilation times can be significantly reduced (e.g., nearly three orders of magnitude [163]). Furthermore, FPGAs – once programmed – are often expected to run for a long time, since context switching incurs a relatively large (seconds long) overhead. Coarsening the units decreases the size of the configurable state, and multiple contexts can be (and are) stored within the same fabric.

QUKU [164]–[166] (Figure 4:a) was one of the earliest CGRA-like overlays on FPGAs, dating back to 2006, and it was actively researched well into 2013. It features a mesh of processing elements, each capable of addition or multiplication, with nearest-neighbor communication and a token-like network. QUKU – as a prototype – was demonstrated on FIR-filters, Sobel, and Laplacian operations, showing improvements over a similar software implementation in the soft-core Microblaze [167] processors or custom circuitry.

Not strictly a CGRA, ZUMA [168] (Figure 4:b) was an early effort to virtualize the fine-grained resources of an FPGA using a "virtual FPGA", for reason of portability, compatibility, and FPGA-like reconfigurability inside of FPGA designs. Similar to a real FPGA, ZUMA discretized the FPGA into logic clusters that contain a crossbar and K-input Look-Up Table with an optional flip-flop capturing the output. Each cluster was connected to a switch box that can be programmed to route the data around. The area cost of using a virtual FPGA could be as low as only 40% more than the barebone FPGA, demonstrating its benefits. Other (even earlier) work was Firm-core virtual FPGA [169], and some more recent efforts include the vFPGA [170].

Intermediate fabrics (IFs) [163] coarsen the FPGA logic by creating a mesh of computational elements of varying sizes, such as for example multipliers and square root functions, where small connectivity boxes (routers) govern the traffic throughout the data-path. IFs were evaluated on image processing (stencil) kernels, and overall showed an on average 17% drop in clock frequency against a gain of $700\times$ in compilation time over using the FPGA alone.

The CREMA [171] overlay architecture was a follow-up of the Butter architecture (described in section IV-B), and targeted both integer and floating-point arithmetic on FPGAs. It was based around a parametrizable VHDL template and supports changing architectural details such as data width and the number of external routing inputs per PE, albeit the size of the CGRA itself was fixed at $4 \times 8$ PEs. CREMA can hold multiple contexts and was evaluated on a matrix-vector multiplication kernel, demonstrating performance improvements of up to four times over the Butter accelerator. The AVATAR [172] CGRA shared many similarities with CREMA but specialized on Fast Fourier Transforms (FFTs) and also had a different layout ($4 \times 16$ vs the $4 \times 8$ in CREMA). The SCREMA [173] CGRA was an extension to CREMA allowing the size of the CGRA to be parametrized, and focused on accelerating matrix-vector multiplications.

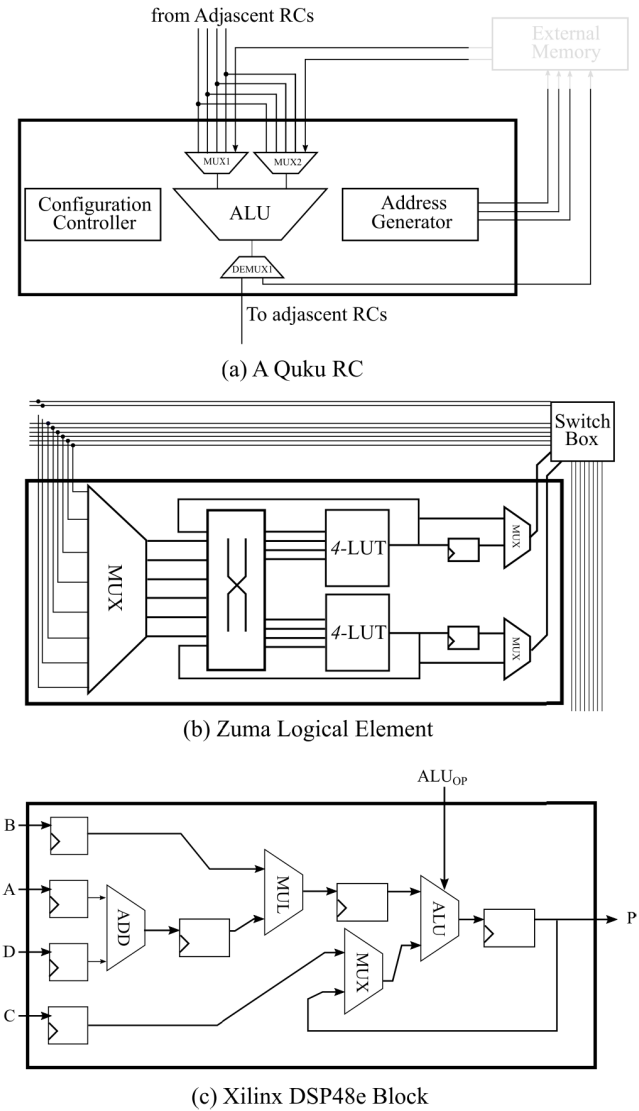

**FIGURE 4.** (a) The QUKU overlay was among the earliest CGRAs running on an FPGA, and adopted a simple scheme where RCs connect to adjacent tiles and (an optional) address generator for reading/writing external memory data. (b) The ZUMA architecture – while not technically a CGRA – was an FPGA-on-FPGA overlay whose RCs mimicked those of the actual FPGA. (c) The Xilinx DSP block is versatile and controllable without reconfiguration, and has been a popular choice as a basis for CGRA-overlays.

The MIN overlay architecture [174] approaches the CGRA design differently; it uses a one-dimensional strip of processing elements whose output is connected to each other through an all-to-all interconnect. Hence, data-flow graphs are cut and fitted onto the linear array, and different parts of the graph are scheduled in time on the array and the interconnect. Different combinations and compositions of the processing elements were evaluated, and the clock frequency, for the most part, ran at 100 MHz, competitive to soft processor cores at the time. Other, arguably less configurable, overlays follow a similar one-dimension strip design, such as the VectorBlox MXP Matrix Processor [175]. The FPCA loop-accelerator described earlier was also prototyped on FPGAs. The READY [176] architecture extends the linear array concept further by also having multiple threads running on the overlay.

The Shared-ALU (SALU) based architecture [177] targets FPGA reconfigurability on two levels: the first level is that a system architect designs the core overlay using a set of FIFO banks, shared ALU clusters, and state machines with microcode. The second (more abstract) level allows dynamic reconfiguration by changing the microcode in the state machines to control how data flows through the array and what operations the shared ALUs use. SALU is (unlike many others) token-driven. The system was evaluated on a Stratix III platform, reaching 204 MHz and demonstrating an FFT use case.

An example of a layered CGRA overlay for FPGAs is the VDR architecture [178]. Here, computational resources are laid out in a one-dimensional strip where each strip is fully-connected to downstream units. Links are unidirectional, and synchronization protocol guides data throughout the data-path. The VDR architecture runs at a clock frequency of 172 MHz, and was shown to be between 3 and 8 times faster compared to the NiosII processor [179] (a well-known soft-core used in FPGA design). Another architecture similar to VDR is the RALU [180].

A flurry of innovative overlays has been introduced since 2012, all centered around the Digital Signal Processing (DSP, Figure 4:c) block of modern FPGAs. The DSP blocks were originally included to allow the use of expensive operation that do not necessarily map well to FPGAs (e.g., multipliers). Since then, DSP blocks have been continuously evolved to include more diverse (various-size multiplication, accumulation, etc.) or more complex (e.g., single-precision floating-point arithmetic [162]) functionality. Some of the vendors (e.g., Xilinx) directly expose the interface to control the different functionality of the DSP blocks to the FPGA fabric, and it was not long before the idea to build CGRA architectures around said DSP blocks emerged. reMORPH [181] was one of the early architectures to adopt this style of reasoning. Several different architectures have been explored around the concept of DSPs, including various topologies (e.g., trees [182]) or adaptation of existing architectures (e.g., DySER using DSP blocks [183]). The strengths of these architectures lie in their near-native performance, where small overlays built around DSPs can run at 390 MHz (or higher).

QuickDough [184], [185] is a design framework for using CGRA overlays on FPGAs, specifically targeting them to assist CPU in accelerating compute-intensive program code. The overlay itself follows the standard layout with a mesh of processing elements, each containing a small instruction memory that sequences the ALU within the processing element. The mesh can interface external memory by enqueuing requests to an address unit. Unique for the architecture is that the two parts (the address unit and the PE mesh) run at two distinctly different frequencies.

Most FPGA overlays presented so far focused exclusively on integer computation. The Mesh-of-FUs [186] was an exception that targeted both integer and floating-point computation. The architecture was similar to other mesh-based approaches, but the work demonstrated high (at the time) performance capabilities of FPGAs also for floating-point operations, reaching nearly 20 GFLOP/s on a Stratix IV [187] device. Using floating-point processing elements seems to incur a 33% area overhead, yielding a smaller CGRA mesh, and also a (arguably negligible) 13% reduction in clock frequency.

A different overlay architecture that targets floating-point operations was the TILT array [188], [189]. The TILT array architecture was conceptually very similar to the MIN overlay. A linear array of processing elements was arranged to communicate with an all-to-all crossbar, which saves the state into an on-chip RAM and relays information to the computation in the next cycle. The authors illustrated the benefits of TILT over High-Level Synthesis (OpenCL) with both comparable performance and improved productivity, reaching an operating frequency of up to 387 MHz on a Stratix V [190].

The URUK [191] architecture took a different approach on how the ALUs inside overlay should be implemented. Rather than having a fixed function, URUK leveraged partial reconfiguration [192], changing the RCs functionality throughout time.

Finally, tools for automatically creating CGRA overlays for FPGAs have emerged, such as the Rapid Overlay Builder [193] and CGRA-ME [61] that simplify generation (and in the case of CGRA-ME also compilation) of applications and overlays.

An interesting observation is that out of the 27 CGRA overlay architectures that we surveyed herein, 17 chose Xilinx FPGAs as the target platform while 10 focused on Intel (then Altera) FPGAs, with some studies using both. There seems to be a favoring of Xilinx architectures, which we believe is due to the more dynamic control that Xilinx offers in their DSP blocks compared to Intel. On the other side, Intel DSPs have (starting from Arria 10 onwards) hardened support for IEEE-754 single-precision floating-point operations, encouraging research on floating-point capable FPGA overlays.

## V. CGRA TRENDS AND CHARACTERISTICS
### A. METHOD AND MATERIALS
For all previous surveyed and summarized work, we collected several metrics associated with each study. These were:

1) **Year** of publication,

2) **Size of the CGRA** array in terms of unique RCs,
3) **Data-path width** of the CGRA (e.g., MATRIX operates on 4-bit while RaPiD operates on 16-bit),
4) **Clock frequency** of operations ($f_{max}$) in MHz as reported in the study,
5) **Power consumption** in Watt. For studies that empirically measured this metric, we collected the `(benchmark, power)` tuple. Otherwise, we used what is reported in the study (often the post place-and-route power estimation),
6) **Technology** (in nm) of architecture when either taped out on silicon, or the standard cell library used with the EDA tools,
7) **Area** ($mm^2$) of the fully synthesized chip as reported in the study. In some cases, we had to manually calculate it based on the individual RC size or based on the gates used (after verification with authors). For FPGAs, we used the chip (BGA) package size and assumed a chip-to-die ratio of 7:1, as has been reported in [194].
8) **Peak performance**, including peak operations per second (OP/s), peak multiply-accumulates per second (MAC/s), Peak floating-point operations per second (FLOPS) as reported in the paper. We differentiate between integer MAC/s and OP/s because some architectures (e.g., EGRA) do not balance them, leading to a large theoretical OP/s but not a proportionally large MAC/s.
9) **Obtained performance** out of the theoretical peak (%). We used what the authors reported. For those cases where authors did not report obtained performance (e.g., only reported absolute time), we derived this metric manually where applicable (otherwise we do not include it), such as for example when the authors report both the input dimension and the execution time (in seconds or cycles) of known applications such as (non-Strassen) matrix-multiplication, FIR-filters, matrix-vector multiplication, etc.

For items 8-9, we ignored studies that only reported relative performance improvements, as it is hard to reason around the performance of a baseline unless explicitly stated. All metrics included have either been directly reported in the seminal publication, have been verified by the authors, or we were confident in our understanding of the architecture to derive them ourselves. We positioned and related our obtained CGRA characteristics against those of modern GPUs. We used NVIDIA GPUs as references with data collected from [195] and integer performance calculated using methods described in [196].

## VI. OVERALL ARCHITECTURAL TRENDS
Figure 5 shows an overview of how CGRAs have changed over time with respect to various metrics. The total number of RCs, as a function of the respective publication year, is shown in Figure 5:a. We see that a majority of CGRAs are quite small (median: 64 RCs) and even smaller for FPGA-based

CGRAs (median: 25 RCs). This is in line with the reasoning that most CGRAs focus on small kernels in the embedded application domain, honoring ILP rather than other forms of parallelism (e.g., thread- or task-level). There are several exceptions to this, such as Garp, which was an early CGRA that used 1/2-bit reconfigurable data-paths and thus needed a large number of RCs to implement various functionality. The other exception is Tartan, where the author's largest evaluated version is up to 25,600 RCs, making it likely the largest CGRA ever simulated; this awe-inspiring size was reached by severely restricting the functionality of the RCs (e.g., there is no multiplication support). Thirdly, the Plasticine architecture can have up to 6208 RCs of varying sorts. Figure 5:b shows the transistor scaling of CGRAs and Nvidia GPUs. As expected, the transistor dimensions have continuously shrunk, as predicted by Moore. Note, however, that both FPGAs and GPUs are (on average) one transistor generation ahead of CGRAs, likely due to most CGRAs being developed by academia and thus restricted to those standard cell libraries available at the time (which usually are not the most recent).

Figure 5:c shows the area of the CGRAs as reported either by the ASIC synthesis tools, estimation by authors, or by the final taped-out chip. We also include the full-size of the FPGA die sizes (that FPGA-based CGRAs have access to), and we position these against the die-size of modern Nvidia GPUs. We can see that the trend of CGRA research is – as with the size of CGRAs – to favor smaller CGRAs, and the median size of the CGRAs is around 13 $mm^2$. Compared to GPUs, which have monotonically increased their size over time, CGRAs have almost done the inverse, and decreased in size. There are two major exceptions: the first is the Imagine architecture, which reported an amazing size of 1000 $mm^2$ (later 144 $mm^2$ in the follow-up paper six years later)– larger than any CGRA or GPU reported to this day. The other larger architecture is the CUDA-programmable SGMF at 800 $mm^2$.

Figure 5:d shows how the reported power consumption of ASIC-based CGRAs has grown over time, and is compared to the Thermal Design Power (TDP) reported for Nvidia GPUs. The CGRAs are experiencing an on-average exponential decrease in power consumption, which is likely due to smaller standard cell libraries coupled with small CGRA size (Figure 5:a,c,d). On the other hand, new generations of Nvidia GPUs tend to draw more power from older generations. The most power-consuming CGRA, out of those reporting, is the Plasticine architecture consuming a maximum of 49 Watt, followed by Raw at 18.2 Watt and IPA at 11.26 Watt. On the opposite side, we find architectures that target bio-signal processing, where power efficiency is critical. Examples include the 167-core processor (60 MHz version) at 99 mW, SYSCORE at 66.3 mW, CMA-1 at 11.2 mW, and CCSOTB2 at 3.45 mW.

Figure 5:e shows how the clock-frequency has changed throughout time for both ASIC and FPGA-based CGRAs, as well as Nvidia GPUs. The GPUs have their clock frequency increased, starting off well-around where CGRAs

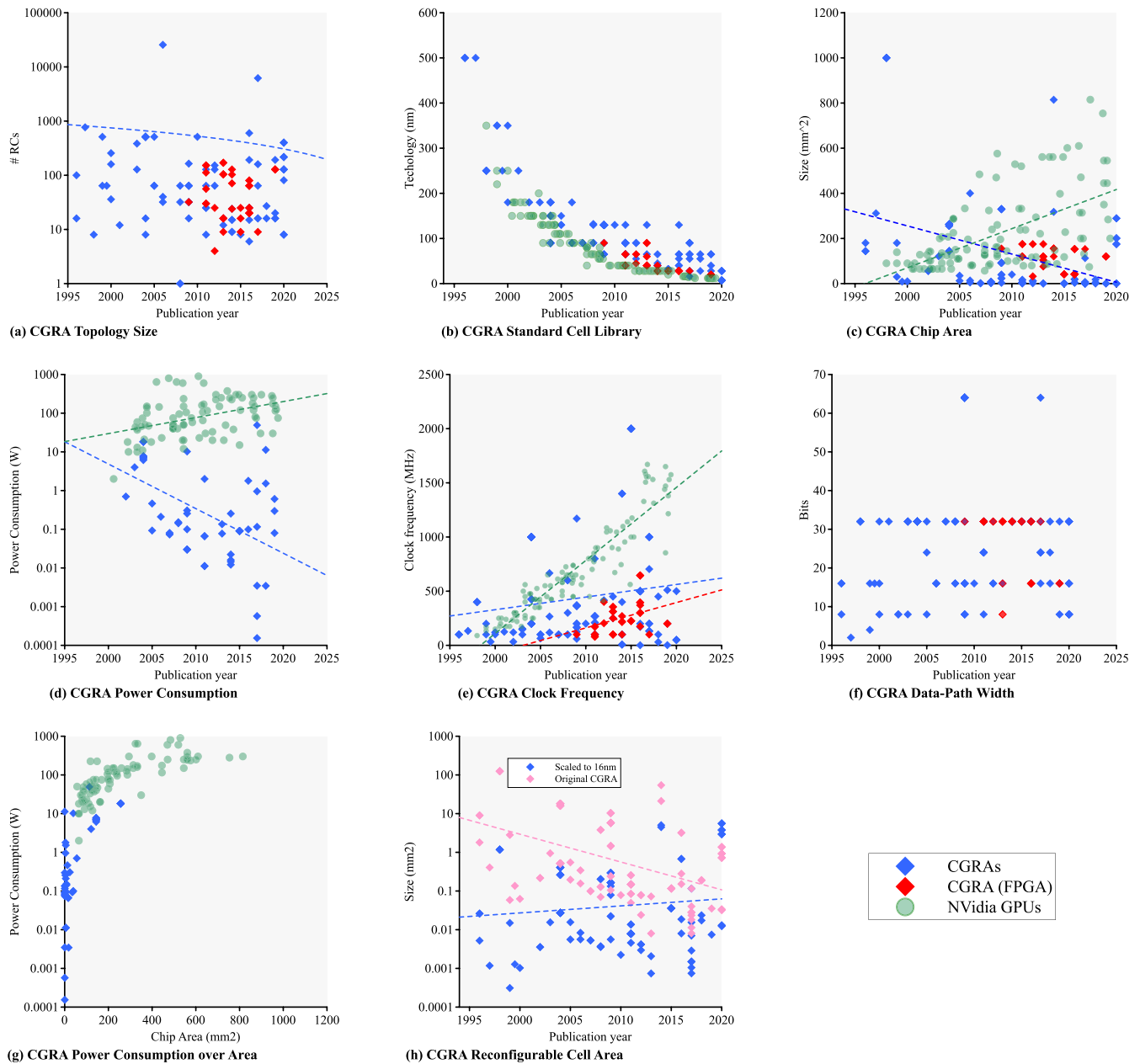

**FIGURE 5.** Architectural trends of CGRAs (both ASIC and FPGAs), showing how different metrics have changed as a function of time, where possible positioned against that of Nvidia GPUs (dotted lines are fit to data to show trend).

operated in the early 2000s to the high-frequency devices we see today. The CGRAs, as one would expect (following the trend of previous graphs), took a different path and instead focused on low-power (and thus low-frequency). The average CGRA will run around 200 MHz, with the frequency slowly increasing as a function of better and smaller standard cell libraries. There are but few CGRAs that operate at high-frequency, and these include the loop-accelerator SEED at 2GHz, the GPU-like SGMF at 1.4 GHz, the 167-core processor (high-frequency version) at 1.17 GHz, and both Plasticine and Wavescalar at 1 GHz. At the opposite edge, we find CGRAs with very low frequency, such as ULP-SRP at (as low as) 7 MHz and the Bio-CGRA [158] at 1 MHz. The FPGA-based CGRAs have less of an opportunity to tune for frequency, as they are often bound by limitation in the fabric itself; however, it is interesting to see that the operation frequency of FPGA-based CGRAs is rivaling most of the ASIC CGRAs.

Figure 5:f shows the chosen data-path width that CGRAs research tends to adopt. Most architecture adopts either a 16-bit (30%) or 32-bit (53%) data-path width; those targeting 16-bit data-path are usually more tailored towards a specific application, such as telecommunication or deep-learning, while those that target 32-bit (or beyond) are

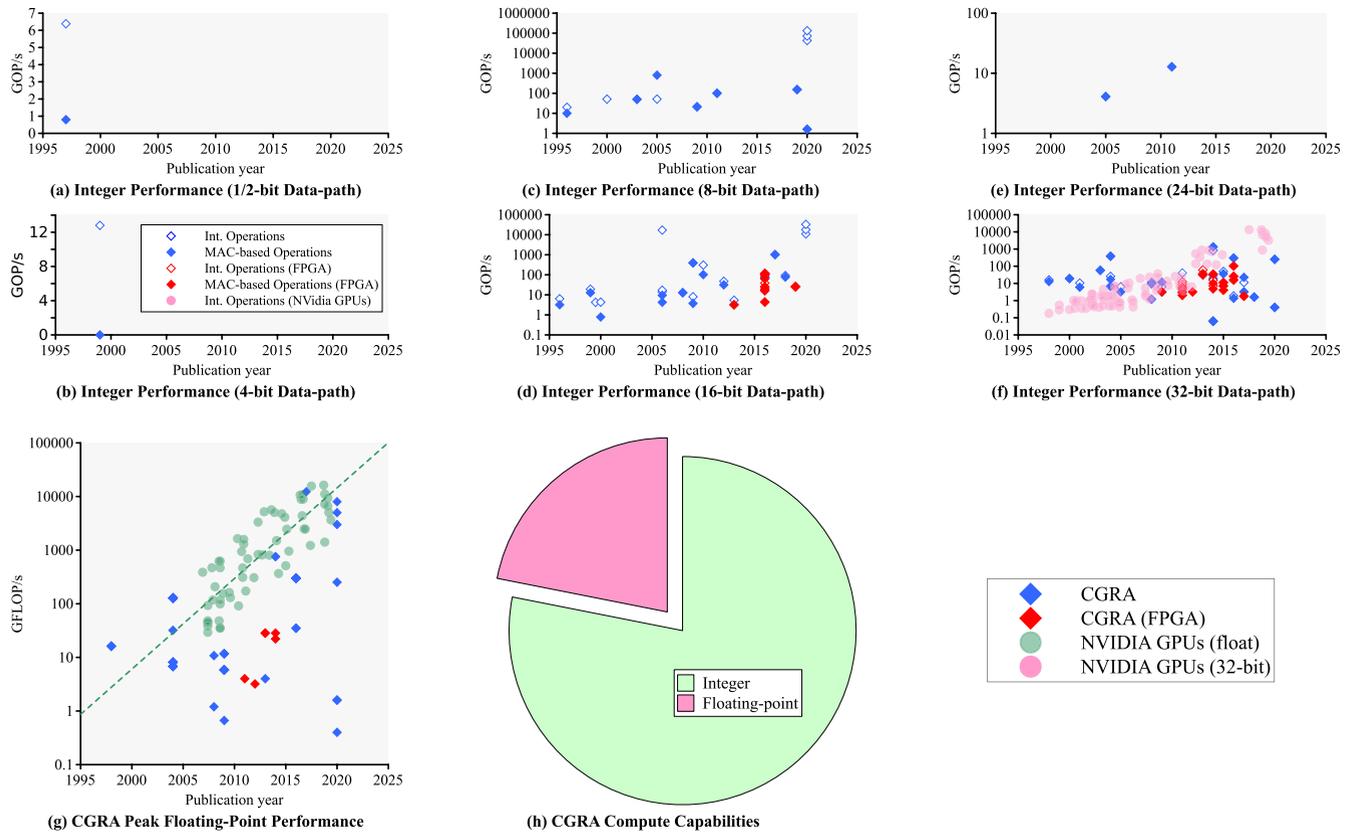

**FIGURE 6.** Integer and floating-point characteristics of CGRAs as they appear in the scientific literature, and compared with historical data of Nvidia GPUs for 32-bit integer and single-precision performance.

more general-purpose. A few (11.5%) target 8-bit architecture, but often have support for chaining 8-bit operations for 16-bit use. MATRIX and Garp target very fine-grained reconfigurability, with 4-bit and 1-/2-bit, respectively. Despite this, we can expect future architectures to include more support for low or hybrid-precision, since it is a reliable way of obtaining more performance while mitigating memory-boundness for applications that permit it.

Figure 5:g shows the power-consumption of CGRAs and GPUs as a function of their respective die sizes. This graph complements the graph in Figure 5:d to show that the low power-consumption of CGRAs is mainly because they are small, with (out of those CGRAs that report both power- and area) only Plasticine coming closer to the trend of GPUs.

Finally, Figure 5:h shows how the individual RC area has grown throughout time, and we see that the size of RCs has been following the technology scaling, and continuously decreased in size. However, when normalizing the CGRAs manufacturing technology to that of 16 nm, we actually noticed a different trend, where the area of individual RCs is increasing, due to incorporating more complex elements (such as wider data-paths, more complex arithmetic units, etc.).

## VII. INTEGER AND FLOATING-POINT PERFORMANCE ANALYSIS

Figure 6 overviews data associated with the pure performance of the CGRAs, often when positioned against that of Nvidia GPUs.

Figure 6:a-f shows the obtained integer performance. Here we distinguish between operations and MAC-based operations in order to reveal architectures that are starved of multipliers. For example, the Tartan CGRA can execute a large number of operations per unit time, but has no support for multiplications, leading to a very low comparable multiply-add performance. Figure 6:a-b show the Garp and MATRIX architecture as the sole candidates for low-precision arithmetic, and that while both of these have comparable high performance (for their time), their multiplication (MAC) performance is lacking (in Garp, the overhead was 32× compared to an addition). Figure 6:c shows 8-bit integer performance, which has recently been of interest to the deep-learning inference community, and where next-generation Xilinx Versal architecture will be capable of reaching thousands of GOP/s of 8-bit integer performance. Figure 6:d shows 16-bit integer performance, showing a continuous growth over the years. Note how the Tartan architecture claims to reach similar performance levels of the upcoming Xilinx Versal CGRA, despite being more than a decade old. Figure 6:e is a special case, since only a few CGRAs (e.g. Cool Mega-Array 1 and SYSCORE) support them. Despite their low visible performance, these devices are actually very power-efficient (see the next section for discussion). Finally, 32-bit integer performance is shown in Figure 6:f, where we also included Nvidia GPU integer performance for comparison. We see that CGRAs have historically been comparable in performance to that of Nvidia GPUs, and even FPGAs are becoming a effective way of obtaining integer performance through CGRAs.

Figure 6:g shows the peak floating-point performance that CGRAs reported over the years. The number of floating-point capable CGRAs prohibits us from drawing any reasonable trend-line, unlike the one for GPUs that exponentially grows with years (together with the die-area, see Figure 5:c). However, those CGRAs that do include floating-point units can compete with the performance of modern GPUs– sometimes even outperform. For example, the Plasticine architecture is capable of delivering 12.3 TFLOP/s of performance, rivaling GPUs from that generation, and the earlier Redefine and SGMF architecture could deliver 300 and 840 GFLOP/s respectively. Even earlier, the WaveScalar architecture was capable of 128 GFLOP/s, which was well ahead of GPUs at that time. At a lower performance, we find architecture such as FloRA (600 MFLOP/s) and the loop-accelerator VEAL (5.4 GFLOP/s). Finally, the Xilinx Versal AI engines will come in different configurations, three shown in the graph (at 3 TFLOP/s, 5 TFLOP/s, and 8 TFLOP/s).

Figure 6:h shows the distribution over the number of CGRAs that supports floating-point versus those that support integer computations. Floating-point support is *clearly underrepresented*, and only one in four architectures support floating-point arithmetics.

## VIII. PERFORMANCE USAGE ANALYSIS

Figure 7:a shows the number of instructions-per-cycle (IPC) that applications/benchmarks experienced when executing on different CGRAs. We see that a majority of CGRAs operate in a fairly low-performance domain, primarily due to their size, and most execute around 12 IPC (median). There are corner cases, such as the Rhyme-Redefine architecture, which aims to explore CGRA in High-Performance Computing, reaching 300+ IPC on selected workloads, or the Deep-Learning SDT-CGRA architecture on inference, reaching 172 IPC. Similarly, Eyeriss is capable of occupying 100% of its resources when inferring AlexNET, yielding astounding 700+ IPC. Most FPGA-based CGRAs also execute less than 100 IPC; this is primarily because the size of most FPGA CGRAs is rather small (see previous Figure 5:a).

Figure 7:b shows the performance that applications experience when running on different CGRA architectures as a function of topology size, where we group CGRA architectures into three groups: small (<16), medium (16-64), and large (>64). As is expected, we see that the performance and obtained IPC grow as the architectures become larger, where applications running on small-sized CGRA

commonly experience (median values) 7.96 IPC, 13.8 IPC on medium-sized CGRAs, and 58.4 IPC on large ones, with outliers being capable of reaching much more. A complementary graph is seen in Figure 7:c, where we see the obtained performance as a fraction of the raw peak performance. The graph reveals that reaching peak-performance becomes harder as the CGRA architecture grows in size, where small architectures commonly reach 25% and more of the peak performance while large architectures reach 18.2%, and medium-sized architecture reach 23.4%. Most of these architectures, as we will discuss later, rely primarily on using software pipelining as their prime source of parallelism, which might not necessarily be able to fill architectures to their maximum.

Figure 7:d shows compute densities and how they have evolved throughout CGRA research history. For comparison, we included Nvidia GPU performance for 32-bit integer and single-precision floating-point. Overall, the compute densities of CGRAs are in-fact comparable to GPUs, and often even surpass. For example, most CGRAs that support 32-bit integer operations consistently pack more compute per $mm^2$ than GPUs. This also holds true for some CGRAs with floating-point support, such as Plasticine. Overall it makes sense that CGRAs pack more compute per unit area, since from an architectural perspective, a CGRA trades much of the silicon used for orchestration (instruction memory, caches, control-planes, etc.) for more compute. The compute densities correlate with the transistor size, and the more recent architectures have a more clear advantage. Note that CGRAs are likely to have an even higher benefit than what the graph shows since the transistor generation usually lags behind one generation compared to commercial GPUs, likely yielding an even higher benefit.

Figure 7:e shows how many operations can be performed per unit power (OPs/Watt) given the architecture. This could arguably be the most important metric, as it serves in part to remedy dark silicon and is also a well-known metric for how power-efficient an architecture/system is (e.g., Green500 [197]). Compared to GPUs, a majority of CGRAs can execute more operations – integer and floating-point alike – per unit power. Some architectures, such as the Plasticine, can be up to two orders of magnitude more power-efficient than GPUs, which offsets any criticism that using TDP as proxy for GPU power consumption may incur. Similar to the reasoning on compute densities, technology scaling impacts power-consumption as well, and in reality it is likely that the power-efficiency is even better than what the graph illustrates.

Figure 7:f-g shows how the area- and power-densities of compute change with respect to frequency. As expected, we see that the CGRAs operate largely in a different region than GPUs– they have high power-efficiency while operating at a lower frequency.

The Nvidia V100 is one of the largest commercial accelerators readily available today, featuring an 815$mm^2$ die built on 12 *nm* technology. The peak performance of the V100 reaches 15.67 TFLOP/s and 13.43 TOP/s (32-bit). We scaled historical CGRAs that reported both area and performance to that of the Nvidia V100 by scaling up the respective chip areas up to the V100's 814 $mm^2$ and also increase the performance of the chip as a function of the increased transistor densities (obtained by moving to 12 *nm*). The transistor densities we used are the same as the one that Nvidia GPUs experienced and can be seen in Figure 8. Albeit this analysis does have several drawbacks, such as power- and thermal-effects (subject to dark silicon [198]) being unaccounted for and frequency un-scaled (we assumed original CGRA frequencies), it does provide a perspective on the question: *what if* CGRAs were on the same playing field as GPUs? The results are seen in of competing with modern GPU designs, assuming we can fully utilize these (potentially over-provisioned) computing resources.

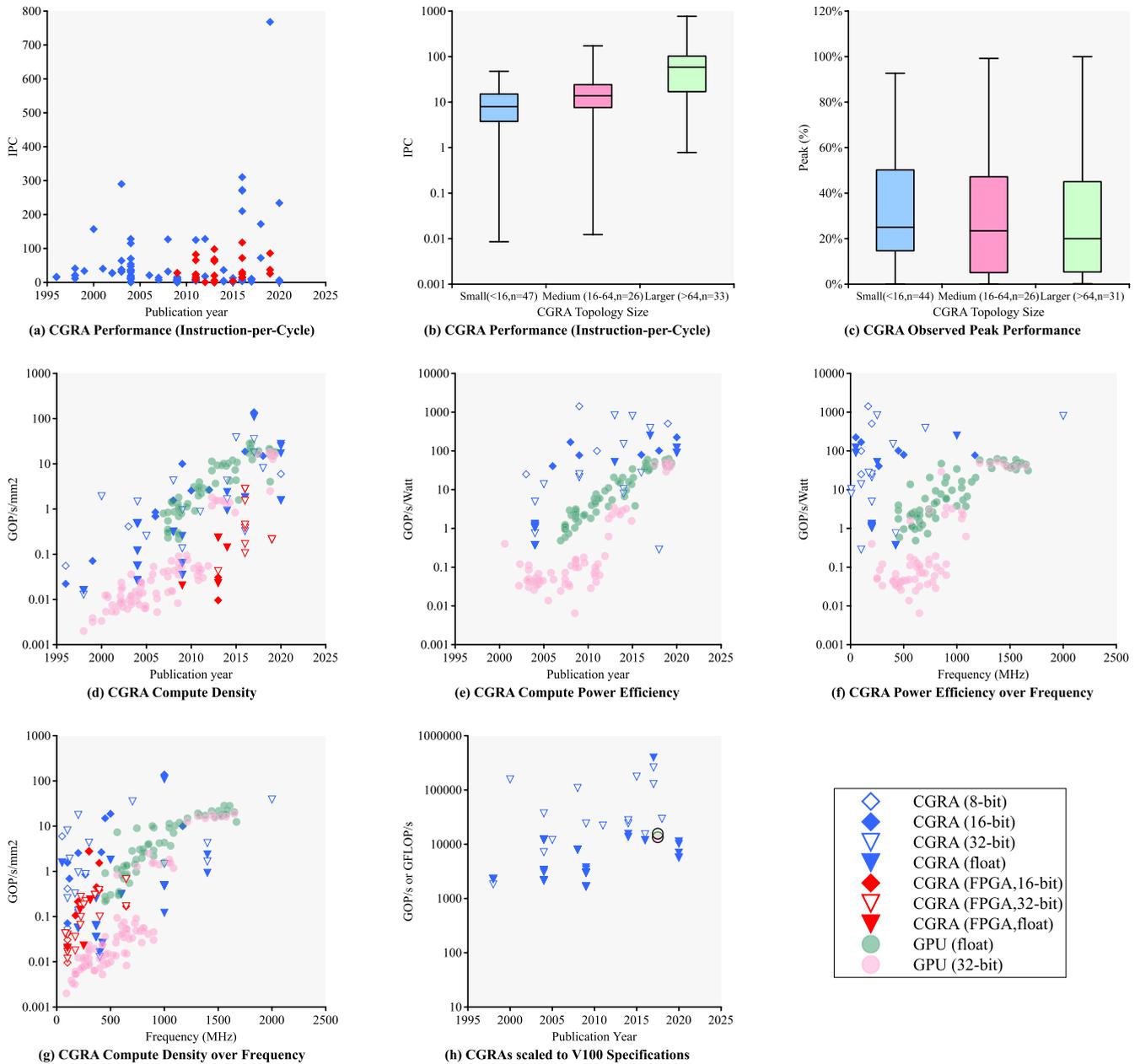

**FIGURE 7.** Performance with respect to time, power, and area of the CGRAs as reported in the scientific literature, compared to that of Nvidia GPUs. The last figure (h) projects the performance of various CGRAs assuming scaling similar to that experienced by the GPUs.

Figure 7:h. We see that while most CGRA architecture would still be below the performance of a V100 (primarily due to the lower clock frequency). However, those architectures whose target indeed is to provide high-performance, such as HyCube, WaveScalar, SGMF, CGRA-ME, and even the older KressArray, show comparable performance to the V100 for either 32-bit integer, floating-point or both. The highest obtained level is demonstrated by Plasticine, which – if given an 815 $mm^2$ design built on 12 nm technology – could reach hundreds of TFLOP/s of performance. While this extrapolation is indeed very limited and ignores many properties, it aims to show that CGRAs have the architectural capability

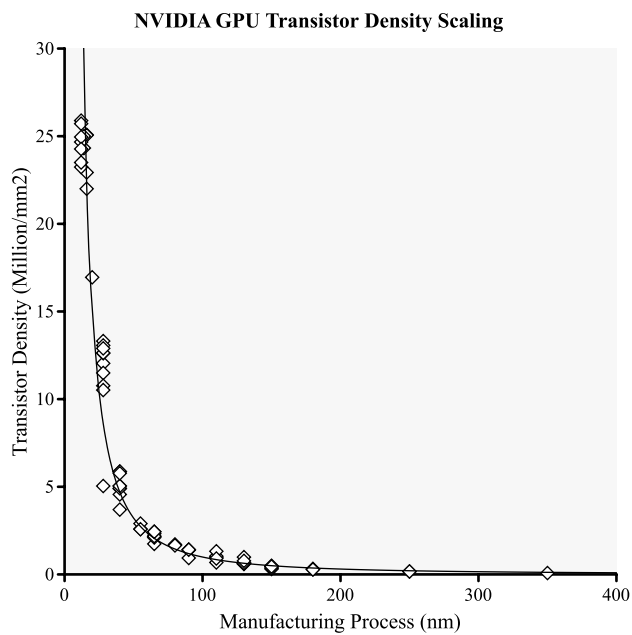

**FIGURE 8.** Transistor-density scaling as experienced by Nvidia GPU chips, which we used to scale existing CGRAs from various times in history.

## IX. DISCUSSION AND CONCLUSION
As we saw in the previous section, a vast majority of CGRAs are fairly small and run at a (comparably) low frequency. This, in turn, leads to very power-efficient designs, allowing CGRAs to be placed into embedded devices such as mobile phones or wearables and operate for hours. This power efficiency, with respect to the performance they provide, allows CGRAs to compete with (and possibly outperform) GPUs, which in turn could lead to a partial remedy for dark silicon. Based on the analysis herein, we argue that CGRAs should be considered a serious competitor to GPUs, particularly in a future post-Moore era when power-efficiency becomes more important.

However, in order to reap the better compute-densities and better power-efficiency that CGRAs offer, larger architectures must be more thoroughly researched. Larger CGRAs, particularly those aimed towards aiding or accelerating general-purpose computing, will be challenging to keep occupied. As we saw in the final graph, CGRAs scaled to the level of an Nvidia V100 will potentially have a peak performance consisting of hundreds of teraflops, but the main question is: will we be able to map and fully utilize all those computing resources for anything but the most trivial kernels?

Several authors have already pointed out that in order to harvest larger CGRAs, we need to complement current ways of extracting instruction parallelism (primary modulo scheduling) with other forms of concurrency. While modern CGRAs (e.g. Plasticine, SIMD-RA) do exploit SIMD-level parallelism, there will, without doubt, be the need to even further research and support programming models such as CUDA (SGMF move towards this direction), multi-threading or even multi-tasking (e.g. OpenMP [199]) should be more aggressively pursued both from an architectural and programmability viewpoint. For example, the recently added task dependencies in frameworks such as OpenMP and OmpSs [200] matches very well to clustered CGRAs that have islands of both compute and scratchpad, where the dependencies would dictate how data would flow on these CGRAs (exploiting both inter- and intra-task level parallelism and data locality).

Another limitation of existing architectures is the application domain on which they accelerate. A large majority of CGRAs target embedded applications such as filters, stencils, decoders, etc. Studies that integrate the CGRA into the back-end of a processor (e.g., TRIPS, DySER) tend to have a more diverse set of benchmarks available, and those studies (e.g., Tartan, SEED, SGMF) that rely only on simulation (without hardware being developed) have the richest set of application support. Despite this, CGRAs suffer from a similar problem that current FPGAs struggle with: we limit our studies to small, simple kernels, rather than studying the impact of these architectures on more complex applications. To give a concrete example, there is no reconfigurable architecture that has seriously considered many of the proxy applications that drive HPC system procurement, such as for example the RIKEN Fiber [201] or ECP benchmark suites [202]. For FPGAs and High-Level Synthesis, this might make sense, since there is always the danger that these large kernels might not fit onto a single FPGA; CGRAs, however, can store multiple contexts and kernels with little overhead in switching between them, opening up possibilities for executing whole applications as well as opportunities to exploit inter-kernel temporal and spatial data locality.

A different challenge with the present (and similar future) surveys lies in the amount of reporting by the different studies. For example, studies that apply a simulation methodology often have a broader benchmark coverage, but fail to report hardware details (e.g., area or RTL-information). At the same time, many CGRAs that were actually implemented in hardware (or RTL) do report area and power-consumption, but limit the benchmark selection and information. This leads to gaps in the graphs, where a high-performance CGRA candidate is represented in one graph (e.g., peak performance), but is absent from another graph (e.g., area), in turn limiting our analysis. This could more clearly be seen in those graphs that use derived metrics, such as performance per power (OPs/W). Similarly, many papers often report relative performance improvement, rather than absolute numbers, leading to difficulty in reasoning around performance across a wide range of CGRAs. We would recommend authors of CGRA papers to, as far as possible, include details on all above listed aspects.

Finally, an important limitation of this study is the tool-agnostic view that we took. Clearly, different CGRAs come with tools of varying maturity, and we must acknowledge that even if certain CGRAs could theoretically operate very well, its performance would be a function of how well the tools (e.g., compiler) did to schedule an application onto them– the tool could even influence the performance more than the CGRA hardware itself. The maturity in tooling infrastructure could also very well explain why some CGRAs were more successful than others. Future work would be advised to (try to) factor out the quality of tools in similar studies to provide an even more accurate view of different CGRAs.

Overall, this survey has shown that there is plenty of room for CGRA research to grow and to continue to be an active research subject for use in future architecture, particularly striving to design high-performance CGRAs that aim at niche or general-purpose computation at scale. As transistor dimensions stop shrinking and Moore's law no longer allows us the architectural freedom of carelessly spending silicon, reconfigurable architectures such as CGRAs might excel at providing performance in a post-Moore era.

## ACKNOWLEDGMENT

The authors would like to thank the anonymous reviewers who have helped improving this survey paper. This article is based on results obtained from a project commissioned by New Energy and Industrial Technology Development Organization (NEDO).

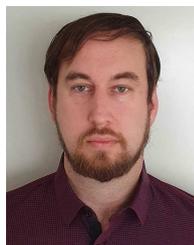

**ARTUR PODOBAS** received the Ph.D. degree from the KTH Royal Institute of Technology, in 2015. From 2015 to 2016, he was a Postdoctoral Fellow of Denmark Technical University, DTU Compute. From 2016 to 2018, he was a Japan Society for the Promotion of Science (JSPS) Fellow of the Global Scientific Information and Computing Center (GSIC), Tokyo Institute of Technology, Japan. He held a postdoctoral position at the Processor Research Team, RIKEN Center for Computational Science, Japan. He is a Researcher with the Department of Computational Science and Technology, KTH Royal Institute of Technology. His research interests include both reconfigurable architectures (FPGAs and CGRAs), neuromorphic computing, parallel architectures, and high-level synthesis for high-performance systems (HPC).

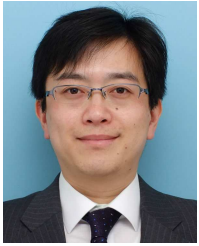

**KENTARO SANO** received the Ph.D. degree from the GSIS, Tohoku University, in 2000. From 2000 to 2005, he was a Research Associate with Tohoku University. From 2005 to 2018, he was an Associate Professor with Tohoku University. He was a Visiting Researcher with the Department of Computing, Imperial College London, in 2006, and Maxeler Corporation, in 2007. Since 2017, he has been a Team Leader of the Processor Research Team with the RIKEN Center for Computational Science (R-CCS). His research interests include FPGA-based high-performance reconfigurable computing systems, especially for scientific numerical simulations and machine learning, high-level synthesis compilers and tools for reconfigurable custom computing machines, and system architectures for next-generation supercomputing based on the data-flow computing model.

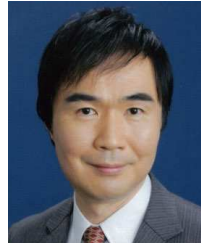

**SATOSHI MATSUOKA** received the Ph.D. degree from The University of Tokyo, in 1993. He has been a Full Professor with the Global Scientific Information and Computing Center (GSIC), Tokyo Institute of Technology, since 2000, and the Director of the joint AIST-Tokyo Tech, the Real World Big Data Computing Open Innovation Laboratory (RWBC-OIL), since 2017, and the RIKEN Center for Computational Science (R-CCS) along with Specially Appointed Professor duty at Tokyo Tech, since 2018. He is the leader of the TSUBAME series of supercomputers that won world's first in power-efficient computing. His various major supercomputing research projects are in areas such as parallel algorithms and programming, resilience, green computing, and the convergence of big data/AI with the HPC. He has written over 500 articles and chaired numerous ACM/IEEE conferences, including the Program Chair of the ACM/IEEE Supercomputing Conference (SC), in 2013. As a Fellow of the ACM and European ISC, he won many awards, including the JSPS Prize from the Japan Society for the Promotion of Science, in 2006, presented by his Highness Prince Akishino, the ACM Gordon Bell Prize, in 2011, the Commendation for Science and Technology by the Ministry of Education, Culture, Sports, Science, and Technology, in 2012, the 2014 IEEE-CS Sidney Fernbach Memorial Award, the highest prestige in the field of HPC, and recently the HPDC Achievement Award from the ACM, in 2018.

. . .